\documentclass[useAMS,usenatbib]{mn2e}
\usepackage{epsfig,rotate,graphicx}
\usepackage[fleqn]{amsmath}
\usepackage{subfigure}
\usepackage{lscape}
\usepackage{bm}

\newcommand{\p}{\partial}

\newcommand{\be}{\begin{equation}}
\newcommand{\ee}{\end{equation}}
\newcommand{\gtrsim}{\;\raisebox{-.8ex}{$\buildrel{\textstyle>}\over\sim$}\;}
\newcommand{\lesssim}{\; \raisebox{-.8ex}{$\buildrel{\textstyle<}\over\sim$}\;}

\newcommand{\bv}{\bm{v}}
\newcommand{\rin}{r_\mathrm{in}}

\newcommand{\ciso}{c_\mathrm{iso}}

\newcommand{\teta}{\tilde{\eta}}

\def\la{\left\langle\rule{0pt}{3em}}
\def\ra{\right\rangle}

\usepackage{array,booktabs,tabularx}
\newcolumntype{R}{>{\centering\arraybackslash}X} 

\title[Gaps in non-isothermal discs]{Gap formation and stability in 
  non-isothermal protoplanetary discs} 

\author[Les and Lin]{Robert Les
  $^1$\thanks{robert.les@mail.utoronto.ca} and Min-Kai Lin $^{1,2}$
  \thanks{ minkailin@email.arizona.edu} \\ 
  $^1$Canadian Institute for Theoretical Astrophysics,  
  60 St. George Street, Toronto, ON, M5S 3H8, Canada \\
  $^2$Department of Astronomy and Steward Observatory, University of
  Arizona, 933 North Cherry Avenue, Tucson, AZ 85721, USA 
}

\begin{document}

\maketitle
\begin{abstract}
  Several observations of transition discs show lopsided
  dust-distributions. A potential explanation is the formation of a
  large-scale vortex acting as a dust-trap at the edge of a gap opened
  by a giant planet. Numerical models of gap-edge vortices have
  so far employed locally isothermal discs in which the temperature profile is held fixed, but the 
  theory of this vortex-forming or `Rossby wave' instability was
  originally developed for adiabatic discs.  
  We generalize the study of planetary gap stability to non-isothermal
  discs using customized numerical simulations of disc-planet
  systems where the planet opens an unstable gap. 
  We include in the energy equation a simple cooling function with
  cooling timescale $t_c=\beta\Omega_k^{-1}$, where $\Omega_k$ is
  the Keplerian frequency, and examine the effect of $\beta$ on the
  stability of gap edges and vortex lifetimes. We find increasing
  $\beta$ lowers the growth rate of non-axisymmetric perturbations, and the
  dominant azimuthal wavenumber $m$ decreases.  
  We find a quasi-steady state consisting of one  
  large-scale, over-dense vortex circulating the outer gap edge, typically  
  lasting $O(10^3)$ orbits. 
    We find vortex lifetimes generally increase with the cooling
    timescale $t_c$ up to an optimal value of $t_c\sim 10$
    orbits, beyond which vortex lifetimes decrease. This 
  non-monotonic dependence is qualitatively consistent with 
  recent studies using strictly isothermal discs that vary the disc
  aspect ratio.  The lifetime and observability of gap-edge 
    vortices in protoplanetary discs is therefore dependent on disc
    thermodynamics. 
\end{abstract}


\section{Introduction}\label{intro}
The interaction between planets and protoplanetary discs plays an
important role in the theory of planet formation and disc 
evolution. Disc-planet interaction may lead to the orbital migration
of protoplanets and modify the structure of
protoplanetary discs  \citep[see][for a recent review]{baruteau13}.

A sufficiently massive planet can open a gap in a 
gaseous protoplanetary disc \citep{pap_lin84,bryden99,crida06,fung14}, 
while low mass planets may also open gaps if the disc viscosity is
small enough \citep{li09,dong11,duffell13}. Support for such disc-planet
interaction have begun to emerge in observations of circumstellar
discs that reveal annular gaps
\citep[e.g.][]{quanz13a,debes13,osorio14}, with possible evidence of
companions within them \citep[e.g.][]{quanz13b,reggiani14}. 

A recent theoretical development in the study of planetary gaps is
their stability. When the disc viscosity is low and/or the planet mass
is large, the presence of potential vorticity (PV, the ratio of
vorticity to surface density) extrema can render planetary gaps
dynamically unstable due to what is now referred to as the `Rossby
wave instability' \citep[RWI,][]{lovelace99,li00}. This 
eventually leads to vortex formation 
\citep{li01,koller03,li05,valborro07}, which can significantly affect
orbital migration of the planet  \citep{ou07,li09,yu10,lin10}. 

Vortex formation at gap edges may also have observable 
consequences. Because disc vortices represent pressure maxima, they are
able to collect dust particles 
\citep{barge95,inaba06,lyra13}. Dust-trapping at gap-edge vortices
have thus been suggested to explain asymmetric dust
distributions observed in several transition discs
\citep[e.g.][]{casassus13,marel13,isella13,fukagawa13,perez14,pinilla15}. 

However, studies of Rossby vortices at planetary gap-edges have 
adopted locally isothermal discs, 
where the disc temperature is a fixed function of
position only \citep[e.g.][]{lyra08,lin11a,zhu14,fu14}. On the other hand, the theory of the RWI was in fact
developed for adiabatic discs \citep{li00}, which permits
heating. 
In adiabatic discs, the relevant quantity for stability
becomes a generalization of the PV that accounts for entropy variations
\citep{lovelace99}.

Gap-opening is associated with planet-induced spiral shocks. In an
isothermal disc, PV-generation across these isothermal shocks leads to
the RWI \citep{koller03,li05,valborro07,lin10}.    
However, if cooling is inefficient and the shock is non-isothermal,
then shock-heating may affect gap stability, since the
relevant quantity is an entropy-modified PV (described below), and
there is entropy-generation across the shocks. 

For example, previous
linear simulations of the RWI found  
that increasing the sound-speed favours instability \citep{li00,lin13}.  
In the context of
planetary gaps, however, the increased temperature may also act to
stabilize the disc by making gap-opening more difficult. It is 
therefore of theoretical interest to clarify the effect of heating and
cooling on the stability of planetary gaps. 

In this work, we extend the study of planetary gap stability against
vortex formation to non-isothermal discs. We include in the fluid energy
equation an one-parameter cooling prescription that allows us to probe
disc thermodynamics ranging from nearly isothermal to nearly
adiabatic.      

This paper is organized as follows. In \S\ref{model} we describe the
equations governing the disc-planet system and initial conditions. Our
numerical approach, including diagnostic measures, are given in
\S\ref{method}. We present results from two sets of numerical
experiments. In \S\ref{linear1} we use disc-planet interaction to set
up discs with gaps, but study their stability without further
influence from the planet. 
We then perform long-term disc-planet simulations to examine the
lifetime of gap-edge vortices in \S\ref{nonlinear},  
as a function of the imposed cooling rate. We conclude and summarize
in \S\ref{summary} with a discussion of important caveats.

\section{Disc-planet models}\label{model}
The system is a two-dimensional (2D) non-self-gravitating gas disc orbiting
a central star of mass $M_*$. We adopt cylindrical
co-ordinates $(r,\phi,z)$ centred on the star. The frame is   
non-rotating. Computational units are such that 
$G=M_*=\mathcal{R}=\mu=1$ where $G$ is the gravitational constant,
$\mathcal{R}$ is the gas constant and $\mu$ is the mean molecular
weight. 

The disc evolution is governed by the standard fluid equations  
\begin{align}\label{3d_gov_eq}
  &\frac{\p\Sigma}{\p t}+\nabla\cdot(\Sigma \bv)=0, \\
  & \frac{\p\bv}{\p t}+\bv\cdot\nabla\bv= -\frac{1}{\rho}\nabla p 
  - \nabla{\Phi} + \bm{f}_\nu,\\
  & \frac{\p e}{\p t} + \nabla\cdot(e\bv) = -p\nabla\cdot\bv +
  \mathcal{H} - \mathcal{C}, 
\end{align}
where $\Sigma$ is the surface density, $\bv = (v_r,v_\phi)$ the fluid
velocity, $p$ is the vertically-integrated pressure, $e=p/(\gamma-1)$ is the energy
per unit area and the adiabatic index $\gamma=1.4$ is assumed
constant. 

The total potential $\Phi$ includes the stellar potential, planet potential
(described below) 
and indirect potentials to account for the non-inertial reference
frame. 
In the momentum equations, $\bm{f}_\nu$ represent viscous forces, 
which includes artificial bulk viscosity to handle shocks, and a
Navier-Stokes viscosity whose magnitude is  
characterized by a constant kinematic viscosity parameter
$\nu$. However, we will be considering effectively inviscid discs by
adopting small values of $\nu$.  

\subsection{Heating and cooling}
In the energy equation, the heating term $\mathcal{H}$ is defined as 
\begin{align}
  \mathcal{H} \equiv Q^+ - Q^+_i\frac{\Sigma}{\Sigma_i}, 
\end{align}
where $Q^+$ represents viscous heating (from both physical and
  artificial viscosity) and subscript $i$ denotes
evaluation at $t=0$. The cooling term $\mathcal{C}$ is defined as
\begin{align}
  \mathcal{C} \equiv \frac{1}{t_c}\left(e -
  e_i\frac{\Sigma}{\Sigma_i}\right),  
\end{align}
where $t_c = \beta\Omega_k^{-1}$ is the cooling time,
$\Omega_k=\sqrt{GM/r^3}$ is the Keplerian frequency and $\beta$ is an
input parameter. This cooling prescription allows one 
to explore the full range of thermodynamic response of the disc in a 
systematic way: $\beta\ll1$ is a locally isothermal disc while
$\beta\gg1$ is an adiabatic disc.

Note that the energy source terms
have been chosen to be absent at $t=0$, allowing the disc to be
initialized close to steady state. The $\mathcal{C}$ function attempts
to restore the initial energy density (and 
therefore temperature) profile. In practice, this is a cooling term at
the gap edge because disc-planet interaction leads to heating.  

\subsection{Disc model and initial condition}
The disc occupies $r\in[r_\mathrm{in}, r_\mathrm{out}]$ and
$\phi\in[0,2\pi]$. The initial disc is axisymmetric with  
surface density profile  
 
\begin{align}\label{initial_density}
   \Sigma(r) = \Sigma_\mathrm{ref}\left(\frac{r}{r_\mathrm{in}}\right)^{-s}
    \left[1 - \sqrt{\frac{r_\mathrm{in}}{r + H_i(\rin)}}\,\right] 
\end{align}
where the power-law index $s=2$, $H(r) = \ciso\Omega_k $ defines the disc scale-height 
where $\ciso=\sqrt{p/\Sigma}$ is the isothermal sound-speed. The disc aspect ratio is defined as $h\equiv H/r$ and initially
$h=0.05$. For a non-self-gravitating disc, the surface density scale
$\Sigma_\mathrm{ref}$ is arbitrary. 

The initial azimuthal velocity $v_{\phi i}$ is set by centrifugal balance with
pressure forces and stellar gravity. For a thin disc, 
$v_{\phi}\simeq r\Omega_k$. The initial radial velocity is
$v_{r}=3\nu/r$, where $\nu = \hat{\nu}\rin^2\Omega_k(\rin)$, and we
adopt $\hat{\nu}= 10^{-9}$, so that $|v_{r}/v_{\phi}|\ll1$ and the initial 
flow is effectively only in the azimuthal direction.  With this value 
of physical viscosity, the only source of heating is through
compression, shock-heating (via artificial viscosity) and the
$\mathcal{C}$ function when $e/\Sigma<e_i/\Sigma_i$. 

\subsection{Planet potential}\label{planet_config}
The planet potential is given by 
\begin{align}
  \Phi_p = -\frac{GM_p}{\sqrt{|\bm{r} - \bm{r}_p|^2 + \epsilon_p^2}},
\end{align}
where $M_p$ is the planet mass and we fix $q\equiv M_p/M_*=10^{-3}$
throughout this work. This corresponds to a Jupiter-mass planet if $M_*=M_{\sun}$. 
The planet's position in the disc 
$\bm{r}_p=(r_p,\phi_p)$  and $\epsilon_p=0.5r_h$ is a softening
length with $r_h=(q/3)^{1/3}r_p$ being the Hill radius.  
The planet is held on a fixed circular orbit with $ r_p = 10\rin$ and $\phi_p=\Omega_k(r_p)t$. 
This also
defines the time unit $P_0\equiv 2\pi/\Omega_k(r_p)$ used to describe results. 

\section{Numerical experiments}\label{method}
The disc-planet system is evolved using the 
\texttt{FARGO-ADSG} code \citep{baruteau08, baruteau08b}. This is a modified version 
of the original \texttt{FARGO} code \citep{masset00a} to include the energy 
equation. The code employs a finite-difference scheme similar 
to the \texttt{ZEUS} code \citep{stone92}, but with a modified azimuthal transport 
algorithm to circumvent the time-step restriction set by the fast rotation speed at the 
inner disc boundary. 
The disc is divided into $(N_r,N_\phi)$ zones in the radial and azimuthal directions, 
respectively. The grid spacing is logarithmic in radius and uniform in azimuth.

\subsection{Cooling prescription}
In this work we only vary one control parameter: the cooling
time. 
The cooling parameter $\beta$ is chosen indirectly  through the parameter
$\tilde{\beta}$ such that 

\begin{align}
  t_c(r_p+x_s) = \beta\Omega_k^{-1}(r_p+x_s) = \tilde{\beta} t_{\mathrm{lib}}(r_p+x_s), 
\end{align}
where $x_s$ is the distance from the planet to its
gap edge, and $t_\mathrm{lib}$ is the time interval between successive
encounters of a fluid element at the gap edge and the planet's
azimuth. That is, we measure the cooling time in units of the time
interval between encounters of a fluid element at the gap edge and the
planet-induced shock. 

Assuming Keplerian orbital frequencies and $x_s\ll r_p$
gives $t_\mathrm{lib}\simeq 4\pi r_p/(3\Omega_{kp} x_s)$, where
$\Omega_{kp} = \Omega_k(r_p)$. Therefore   
\begin{align}\label{betatilde}
  \beta = \tilde{\beta} \frac{4\pi r_p}{3x_s} \left(1  - \frac{3x_s}{2r_p}\right), 
\end{align}
where $x_s\ll r_p$ was used again. We use $x_s = 2r_h$ in
Eq. \ref{betatilde}. For a planet mass with $q=10^{-3}$,
Eq. \ref{betatilde} then gives $\beta \simeq 23.9\tilde{\beta}$. In
terms of planetary orbital periods, this is
\begin{align} 
  t_c(r) = \frac{\beta}{2\pi}\left(\frac{r}{r_p}\right)^{3/2}P_0\simeq
  3.8 \tilde{\beta}\left(\frac{r}{r_p}\right)^{3/2}P_0. 
\end{align}

\subsection{Diagnostic measures}

\subsubsection{Generalised potential vorticity}

The generalised potential vorticity is defined as
\begin{align}
  \tilde{\eta} = \frac{\kappa^2}{2\Omega\Sigma}\times S^{-2/\gamma}, 
\end{align}
where $\kappa^2 = r^{-3}\partial_r(r^4\Omega^2)$ is the square of the
epicyclic frequency, $\Omega=v_\phi/r$ is the angular speed, and
$S\equiv p/\Sigma^\gamma$ is the entropy. The first factor is the
usual potential vorticity (PV, or vortensity). 

The generalised PV appears in the description of the linear stability
of radially-structured adiabatic discs \citep{lovelace99,li00}, where
the authors show an extremum in $\teta$ may lead to a dynamical
instability, the RWI. In a barotropic disc where $p=p(\Sigma)$, the entropy factor is 
absent and the important quantity is the PV. 

\subsubsection{Fourier modes} 
The RWI is characterized by exponentially
growing perturbations. Though in this paper we do not consider a
formal linear instability calculation, modal analysis will be useful
to analyse the growth of perturbations with different azimuthal
wavenumbers, which is associated with the number of vortices initially
formed by the RWI.    

The Fourier transform of the time-dependent surface density is
\begin{align}\label{fouriertransform}
  \Sigma_m(r,t) = \int_{0}^{2\pi}
  \Sigma(r,\phi,t) \, \mathrm{e}^{-\mathrm{i}m\phi} \, \mathrm{d}\phi 
\end{align} 
where $m$ is the azimuthal wave number. We define the growth rate
$\sigma$ of the $m^\mathrm{th}$ component of the surface density
through 
\begin{align}\label{growth}
  \frac{d \langle|\Sigma_m|\rangle_r }{dt}= \sigma \langle|\Sigma_m|\rangle_r 
\end{align}
where 
$\langle|\cdot|\rangle_r$ denotes the average of the absolute value
over a radial region of interest. By using Eq.~\ref{growth} the growth
rates of the unstable modes can be found from successive spatial
Fourier transforms over an appropriate period of time. 

\subsubsection{Rossby number}
The Rossby number
\begin{align}
  Ro = \frac{{\bf \hat{z}} \cdot \nabla \times \bv - \langle
    {\bf\hat{z}} \cdot \nabla \times \bv
    \rangle_{\phi}}{2\langle\Omega\rangle_{\phi}},  
\end{align}
is a dimensionless measure of relative vorticity. 
 Here $\langle \cdot \rangle_{\phi}$ denotes an azimuthal
average. Values of $Ro<0$ correspond to anti-cyclonic rotation with
respect to the background shear and thus can be used to identify
vortices and quantify its intensity. 

\section{Growth of non-axisymmetric modes without the influence of the
  planet}\label{linear1} 
In this section, the planet is introduced at $t=20P_0$ and 
its potential is switched on over $10P_0$. At $t=30P_0$ we switch off the
planet potential and azimuthally average the surface density, energy
and velocity fields. (At this point the planet has carved a partial
gap and the RWI has not yet occurred.) Effectively, we initialise the
disc with a gap profile. 
We then perturb the surface density in the outer disc ($r>r_p$) and continue to 
evolve the disc. We impose  sinusoidal perturbations with 
azimuthal wavenumbers $m\in[1,10]$ and random amplitudes within $\pm 0.01$
times the local surface density. 
This procedure allows us to analyse the growth of 
non-axisymmetric modes associated with the gap, but without
complications from non-axisymmetry arising directly from disc-planet
interaction (i.e. planet-induced wakes). 

Note that these `planet-off' simulations are not linear stability
calculations because the cooling term in our energy equation
restores the initial temperature profile corresponding to constant
$H/r=0.05$, rather than the heated gap edge. However, we will 
examine a nearly adiabatic simulation in \S\ref{adiabatic_section},
which is closer to a proper linear problem. 

Simulations here employ a resolution of $(N_r,N_{\phi})=(1024,2048)$
with open boundaries at $r=r_\mathrm{in}$ and
$r_\mathrm{out}=25r_\mathrm{in}$. We compare cases with
$\tilde{\beta}=0.1,1,10$ corresponding to fast, moderately, 
and slowly cooled discs. 

\subsection{Gap structure}
\begin{figure}
  \includegraphics[width=\linewidth,clip=true,trim=0.5cm
    2cm 0cm 0cm]{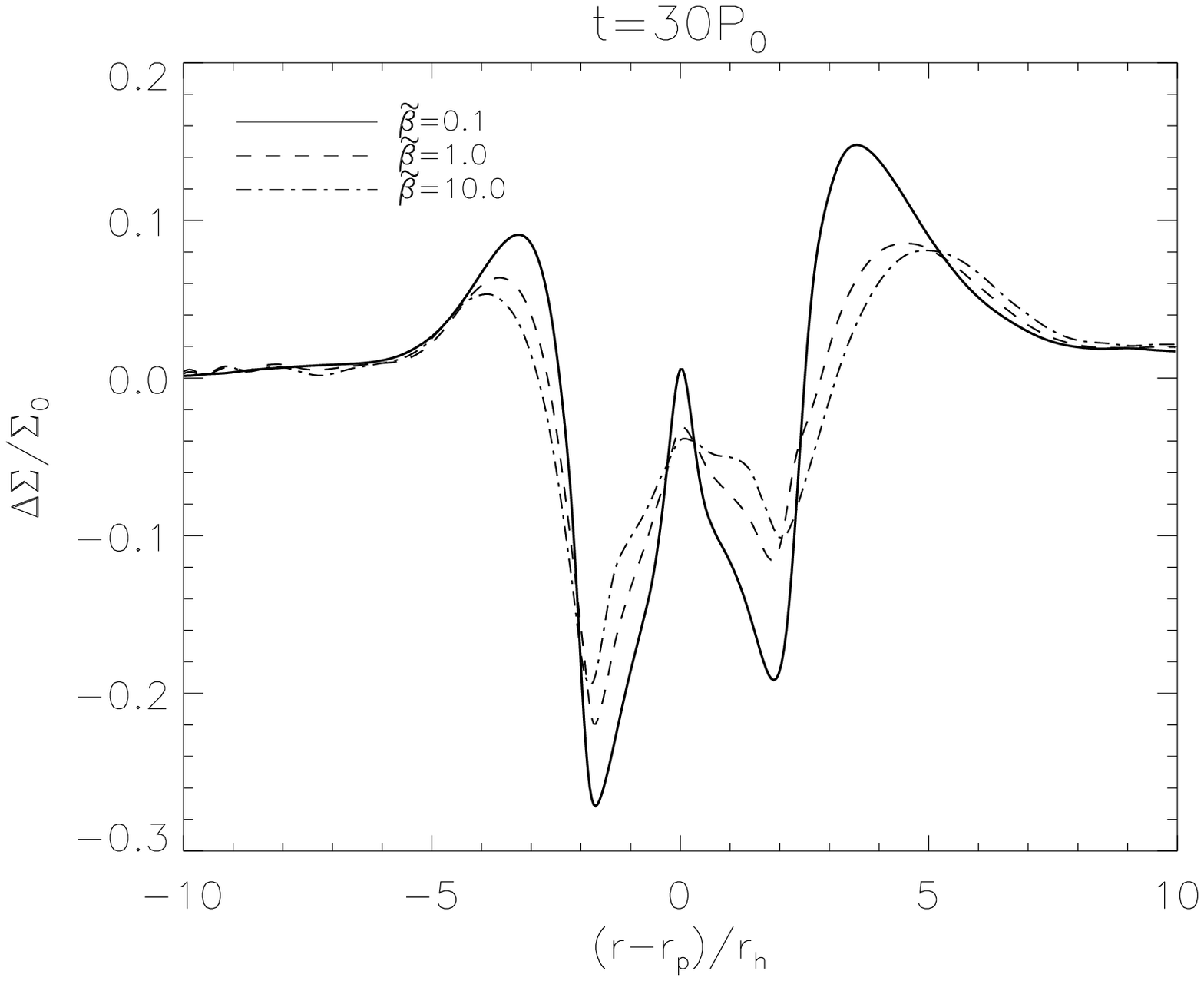}
  \includegraphics[width=\linewidth,clip=true,trim=0.5cm
    2cm 0cm 1cm]{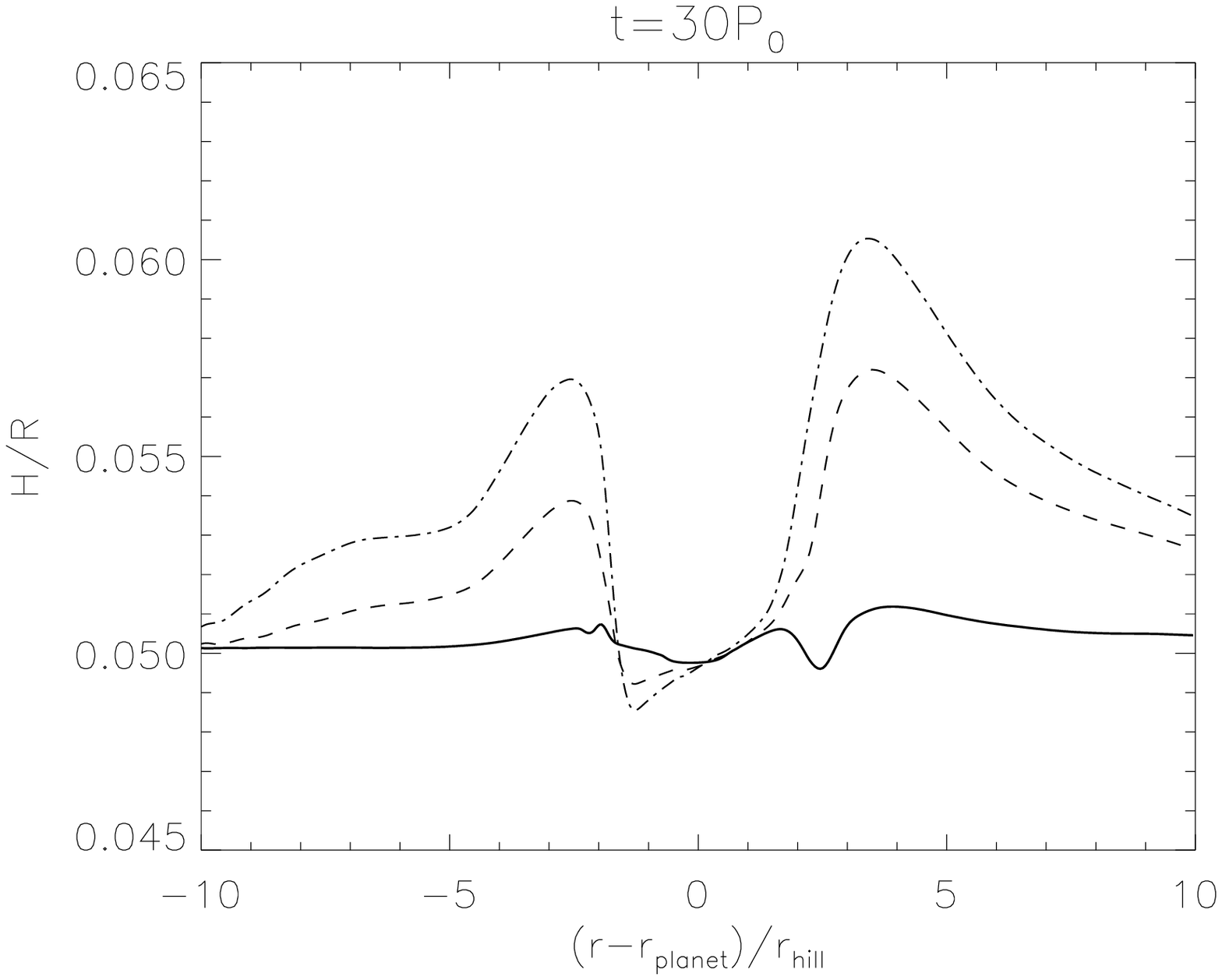}
\includegraphics[width=\linewidth,clip=true,trim=0.5cm
    0.5cm 0cm 1cm]{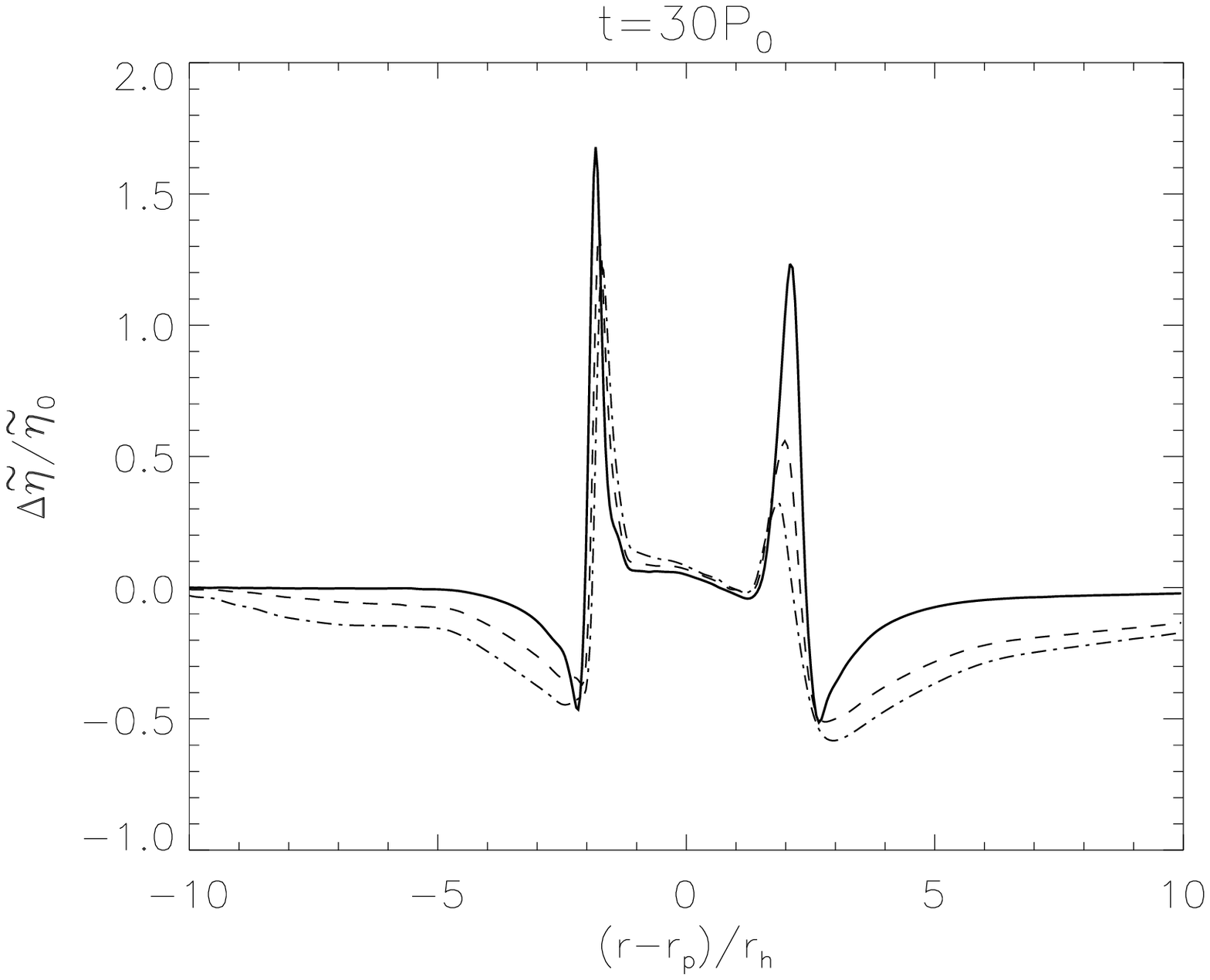}
  \caption{Azimuthally averaged gap profiles at $t=30P_0$ for the
    initial partial gap opened before instability emerges for fast
    (solid, $\tilde{\beta}=0.1$), moderate
    (dashed,  $\tilde{\beta}=1$), and slow cooling (dashed-dot,
    $\tilde{\beta}=10$). The relative surface density 
    perturbation (top), disc aspect ratio (middle) and generalised
    vortensity perturbation (bottom) are shown. \label{intial1D}}  
\end{figure}

We first compare the gap structures formed by planet-disc
interaction as a function of the cooling time. The azimuthally-averaged 
gap profiles are shown in Fig. \ref{intial1D} for different values of 
$\tilde\beta$. Gaps formed with lower $\tilde\beta$ (faster cooling) 
are deeper with steeper gradients at the gap edges. Faster cooling rates also 
increase (decrease) the surface density maxima (minima). However, a
clean gap does not form in this short time period.%

Increasing $\tilde\beta$ leads to higher disc aspect ratios $h=H/r$,
i.e. higher temperatures. Heating mostly occurs at the gap edges
due to planet-induced spiral shocks. Increasing the cooling timescale 
implies that this heat is retained in the disc. In the inviscid limit the gap
opening condition is $r_h\gtrsim H$ or $q\gtrsim 3h^3$
\citep{crida06}, which indicates that for hotter discs (higher
$h$), it becomes more difficult for a planet of fixed $q$ to open a
gap. This explains the shallower gaps in surface density when
$\tilde{\beta}$ is increased.

The important consequence of a heated gap edge is that the
generalised vortensity profiles, $\tilde{\eta}$, becomes smoother with increasing
cooling times, with the extrema becoming less pronounced. Previous locally
isothermal disc-planet simulations show the RWI associated with PV
minima \citep{li05,lin10}. We can therefore expect the RWI to be associated with
minima in the generalised vortensity (corresponding to local surface
density maxima) in the non-isothermal case. Because the extrema are
less sharp, the RWI is expected to be weaker and the gap to be more
stable with longer cooling times.

However, we remark that the change in the gap structure becomes less 
significant at long cooling times, as Fig. \ref{intial1D} shows that
the profiles with $\tilde{\beta}=1$ and $\tilde{\beta}=10$ are 
similar. This implies that the effect of cooling timescale on the
RWI through the set up of the gap profile, becomes less important for
large $\tilde{\beta}$. 

\begin{table}
  \centering
  \caption{Dominant mode and growth rates for
    $\tilde{\beta}=0.1,1.0,10.0$ (fast, moderate, and slow cooling)
    values during `planet-off' simulations \label{modetable}} 
  \hfill
  \begin{minipage}{0.3\linewidth}
    \begin{tabularx}{\textwidth}{l R} 
      \multicolumn{2}{c}{$\tilde{\beta}=0.1$} \\ 
      \toprule
      $m$ & $10^2\sigma/\Omega(r_p)$ \\
      \midrule
      6 & 7.3 \\
      7 & 7.8 \\
      8 & 7.9 \\
      9 & 7.9 \\
      10 & 6.8 \\ 
      \bottomrule
    \end{tabularx}
  \end{minipage}
  \hfill
  \begin{minipage}{0.3\linewidth}
    \begin{tabularx}{\textwidth}{l R} 
      \multicolumn{2}{c}{$\tilde{\beta}=1.0$} \\ 
      \toprule
      $m$ & $10^2\sigma/\Omega(r_p)$ \\
      \midrule
      3 & 2.0 \\
      4 & 2.2 \\
      5 & 2.3 \\
      6 & 1.6 \\
      7 & 1.1 \\ 
      \bottomrule
    \end{tabularx}
  \end{minipage}
  \hfill
  \begin{minipage}{0.3\linewidth}
    \begin{tabularx}{\textwidth}{l R} 
      \multicolumn{2}{c}{$\tilde{\beta}=10.0$} \\ 
      \toprule
      $m$ & $10^2\sigma/\Omega(r_p)$ \\
      \midrule
      1 & 1.1 \\
      2 & 1.6 \\
      3 & 1.7 \\
      4 & 1.2 \\
      5 & 0.1 \\ 
      \bottomrule
    \end{tabularx}
  \end{minipage}
  \hfill
\end{table}

\subsection{Axisymmetric stability}
The initial planet-disc interaction form bumps and grooves in the gap profiles
which can potentially be unstable due to axisymmetric instabilities. The
generalised local axisymmetric stability condition is the Solberg-Hoiland
criterion,  
\begin{align}
  \kappa^2+N^2 \geq 0 
\end{align}
where
\begin{align}
 N^2=\frac{1}{\Sigma} \frac{\partial P}{\partial r}
 \left(\frac{1}{\Sigma} \frac{\partial \Sigma}{\partial
     r}-\frac{1}{\gamma P} \frac{\partial P}{\partial r}  \right) 
\end{align}
is the square of the Brunt-V\"ais\"al\"a frequency. 
At the outer gap edge $r=r_p+2.5r_h$,  where the RWI is excited
(see below),  we find $\kappa^2 + N^2$ reaches local minimum with a value
$\sim 0.45 \, \Omega^2(r_p)$
 for all $\tilde\beta$. The
Brunt-V\"ais\"al\"a frequency at the outer gap edge is
$N\sim 0.1 \, \Omega(r_p)$,  
decreasing marginally with longer cooling rate. 
The Solberg-Hoiland criteria is similarly satisfied for the entire 2D disc throughout the 
simulations.
Thus for all values of $\tilde\beta$ the planet-induced gaps are
stable to axisymmertic perturbations. 

\subsection{Non-axisymmetric instability}\label{linear}
\begin{figure}
  \includegraphics[width=\linewidth,clip=true,trim=1.2cm
  0cm 0cm 0cm]{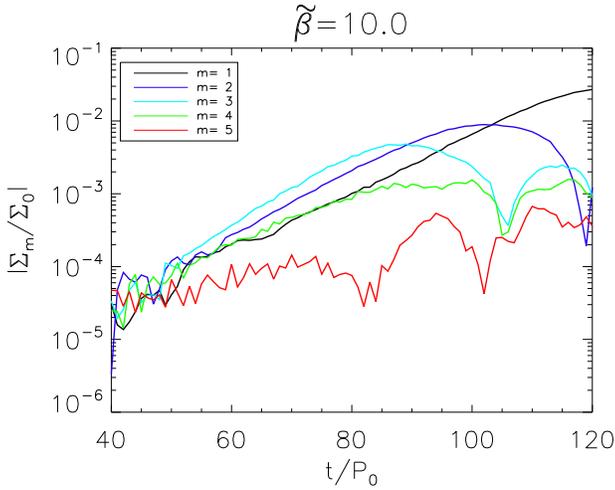}
  \caption{Evolution of azimuthal Fourier modes of disc surface
    density, non-dimensioqnalized by the initial axisymmetric component 
    $\Sigma_0(t=0)$, for the
    `planet-off' simulation with $\tilde{\beta}=10$. Colours correspond
    to different $m$ values. The $m=3$ component is the fastest growing
    mode during linear growth with a growth rate of 
    $\gamma=0.017\Omega(r_p)$.\label{linearmodes}}
\end{figure}

We now examine the evolution of the gap for $t>30P_0$, with the
planet potential switched off, but with an added surface density
perturbation. For all three cooling times $\tilde{\beta}=0.1,\, 1.0,\,
10$, we observe exponential growth of non-axisymmetric
structures. An example is shown in Fig. \ref{linearmodes} for 
$\tilde{\beta}=10$. We characterize these
modes with an azimuthal wavenumber $m$ and growth rate $\sigma(m)$ as defined by
Eq.~\ref{fouriertransform}---\ref{growth}. Mode amplitudes were
averaged over $r-r_p\in[2,5]r_h$. Table \ref{modetable}
lists the growth rates measured during 
linear growth for 5 values of $m$ centred around that with maximum
growth rate. 

\begin{figure}
  \centering
  \subfigure{
    \includegraphics[width=0.3\linewidth]{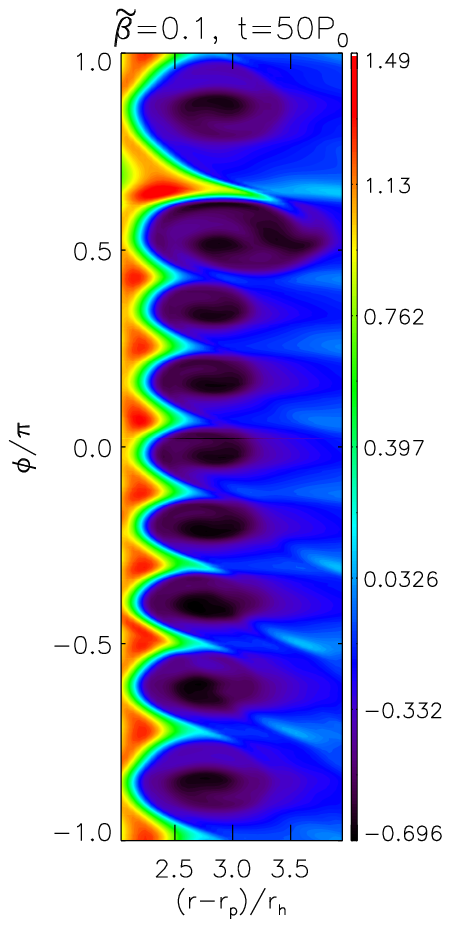}
  }
\hfill
  \subfigure{
    \includegraphics[width=0.3\linewidth]{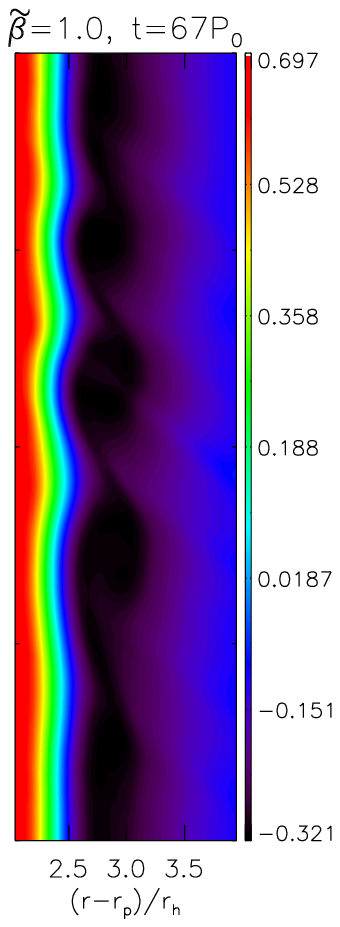}
  }
\hfill
  \subfigure{
    \includegraphics[width=0.3\linewidth]{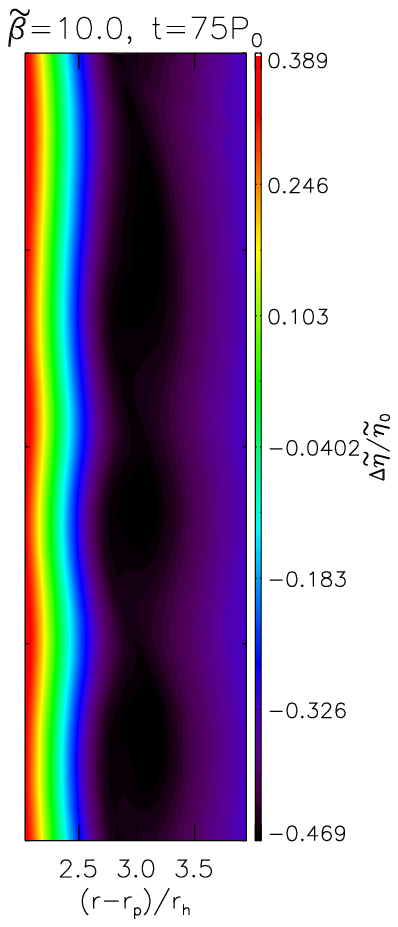}
  }  
  \caption{Generalised vortensity perturbation (relative to $t=0$) for
    cases of $\tilde{\beta}=0.1,1,10$ (left,middle,right) during
    the growth of non-axisymmetric modes. The planet potential has
    been switched off.  The number of vortices
    decrease as $\tilde{\beta}$ increases. Note that snapshots are
    taken later for increasing $\tilde{\beta}$ because it takes longer
    for the vortices to grow and become visible with increasing cooling time. 
    \label{2Dlinear} 
  } 
\end{figure}

Table \ref{modetable} show that as
$\tilde{\beta}$ is increased from $ 0.1\rightarrow10$ the dominant
azimuthal Fourier mode decreases from $ m=9\rightarrow3$ and the
respective growth rate decreases from $ \gamma/\Omega(r_p)=0.079
\rightarrow 0.017$. However, despite two orders of magnitude increase in the
cooling time, the instability remains dynamical with characteristic  growth time
$\lesssim 10P_0$. Snapshots of the instability in for  
the different $\tilde\beta$ are shown in Fig \ref{2Dlinear}. 


These `planet-off' simulations show that gap edges become more stable with
longer cooling times. This is expected because larger $\tilde{\beta}$
results in hotter gap profiles at $t=30P_0$ with less pronounced
generalised vortensity minima. Stabilization with increased
cooling time is therefore due to a smoother basic state for the
instability, as it is more difficult for the planet to open a gap if
the disc is allowed to heat up. 

 For completeness we also simulated a locally
  isothermal disc with $h=0.05$ where the
  sound-speed is kept strictly equal to its initial value.  This
  simulation yield a most unstable growth rate
  $\sigma/\Omega(r_p)\simeq 0.05$ at $m=5$,   
  compared with a value of $0.085$ at $m=6$ for a corresponding
  simulation that includes the energy equation but with rapid cooling
  $\tilde{\beta}=0.01$. However, the vortex evolution is similar. 

\subsubsection{Nearly adiabatic discs}
\label{adiabatic_section}

The above `planet-off' simulations are not formally linear
stability calculations, because the cooling time is comparable or shorter
than the instability growth time, $t_c\lesssim\gamma^{-1}$.  
Thus the disc cools back to its initial temperature corresponding to
$h=0.05$ before or during the instability growth, so we do not 
have a steady basic state to formulate a standard linear stability 
problem. 

In order to capture the effect of a heated gap edge, we ran a simulation with 
$\tilde{\beta}=100$, corresponding to an almost adiabatic disc.  
In this simulation the cooling rate is slow enough that the gap 
temperature profile (e.g. middle panel of Fig. \ref{intial1D}) changes
only marginally over the instability growth timescale. 

We find very similar gap profiles and mode growth rates for
$\tilde{\beta}=100$ as with $\tilde{\beta}=10$. At $t=30P_0$, the disc only heats up to
values $h\simeq0.06$ in the nearly adiabatic case. This is close to
the original temperature of $h=0.05$, so linear growth rates are not expected
to change significantly \citep{li00}. 

According to \cite{li00}, increasing $h$ increases linear growth rates
of the RWI because it is pressure-driven. However, in the case 
of disc-planet interaction, increasing $h$ has a stabilizing effect
through the setting up the gap profile because it results in smoother gap
edges. The fact that we observe smaller growth rates as $h$ is
increased indicates that for planetary gaps, the importance of $h$ on
the \emph{linear} RWI is through setting up the gap profile, i.e. basic
state for the instability (as opposed to the linear response). 

\subsection{Long term evolution} \label{nonlinearplanetoff}

We also extended these `planet-off' simulations into the non-linear
regime. After the linear growth phase of the vortices, vortex merging
takes hold on timescales of up to $150P_0$, until there is one vortex
left. We find the vortex merging
time is dependent on the growth rates of the modes and saturation
timescales, with the slowest growing modes in $\tilde\beta=10$ taking
the longest to merge.  

\begin{figure}
  \includegraphics[width=\linewidth,clip=true,trim=0.5cm
  0cm 0cm 1.1cm]{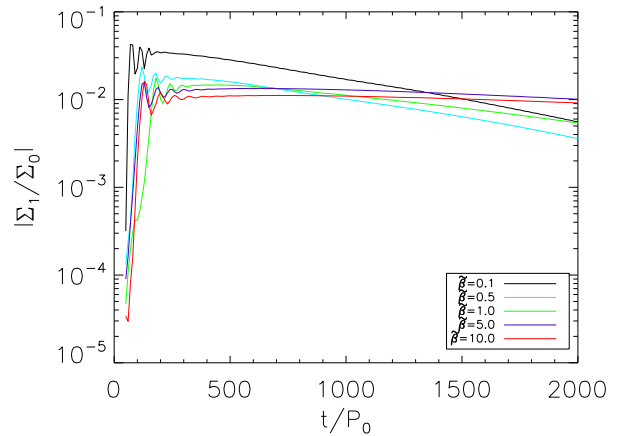}
  \caption{Long term simulations without the planet potential after
    the gap is set up. The $m=1$ surface density component,
    non-dimensionalized by the initial $m=0$ component, at the
    outer gap edge is shown as a function of the cooling timescale. 
  } \label{planetofflifetimeplot}
\end{figure}

Fig.\ref{planetofflifetimeplot} shows evolution of the $m=1$ surface
density component, which represents the amplitude of the post-merger
single vortex. For completeness we also ran intermediate cases with
$\tilde{\beta}=0.5$ and $5.0$. The amplitude of the initially formed vortex was 
found to decrease with increased cooling rate. The vortices simply decay on a
timescale of $O(10^3)$ orbits with faster decay for stronger vortices
(which are obtained with faster cooling rates). 

This decay is probably due to numerical viscosity. During the
  slow decay the vortex elongates (weakens) while its radial width
  remains $O(H)$, so its surface density decreases. In addition, for 
  $\tilde{\beta}=0.1,\,0.5$ and $1.0$ we also observe the appearance of
  spiral waves associated with the vortex, which may contribute to
  its dissipation (see below). 
We will see in the
next section that this decay after linear growth is very
different to when the planet potential is kept on.

\section{Non-linear evolution of
  gap-edge vortices with finite cooling time}\label{nonlinear} 
We now examine long-term simulations of gap-edge vortices for
$\tilde{\beta}=0.1,0.5,1,5,10$. (Additional cases are presented in 
  \S\ref{lifetime_discuss} when examining vortex lifetimes as a
  function of $\tilde{\beta}$.)  The planet potential is kept on
throughout. We employ a grid with $(N_r,N_{\phi})=(512,1024)$ in order
for these 
simulations to be computationally feasible. We also use a larger
disc with $r_{\mathrm{out}}=45r_{\mathrm{in}}$ to minimise boundary
effects on vortex evolution, and apply open boundaries at
$r=r_\mathrm{in},\,r_\mathrm{out}$. 

We comment that lower-resolution simulations with
$(N_r,N_{\phi})=(256,512)$ show similar behaviour and trends as the
high-resolution runs reported below. 

\subsection{Generic evolution} 
\begin{figure}
  \includegraphics[width=\linewidth,clip=true,trim=0.5cm
  0cm 0cm 1cm]{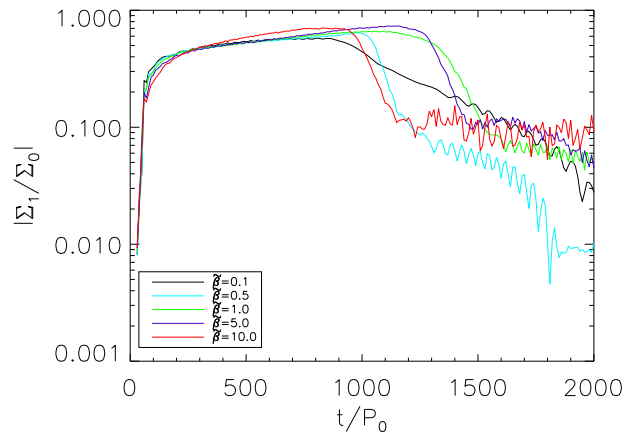}
  \caption{Evolution of the non-dimesnionalized $m=1$ surface density component
    for long term
    simulations with the planet potential kept on. Notice there is
    vortex growth after the initial linear growth, which
    contrasts to Fig. \ref{planetofflifetimeplot}, where vortices
    decay in the absence of continuous disc-planet interaction. 
    \label{lifetimeplot}}   
\end{figure}

The linear growth of the RWI and vortex-formation is followed by 
vortex merging. We now find merging timescales independent of
$\tilde\beta$, and by $60P_o$ only one vortex remains. 
The evolution of the amplitude of the $m=1$ surface density component,
averaged over $r-r_p\in[2,10]r_h$, is shown in Fig.~\ref{lifetimeplot} 
for different $\tilde\beta$. The initial, post-merger vortex amplitude
is found to be weaker for longer cooling rates (which have smaller 
linear growth rates). 

\begin{figure}
  \includegraphics[width=\linewidth,height=\linewidth]{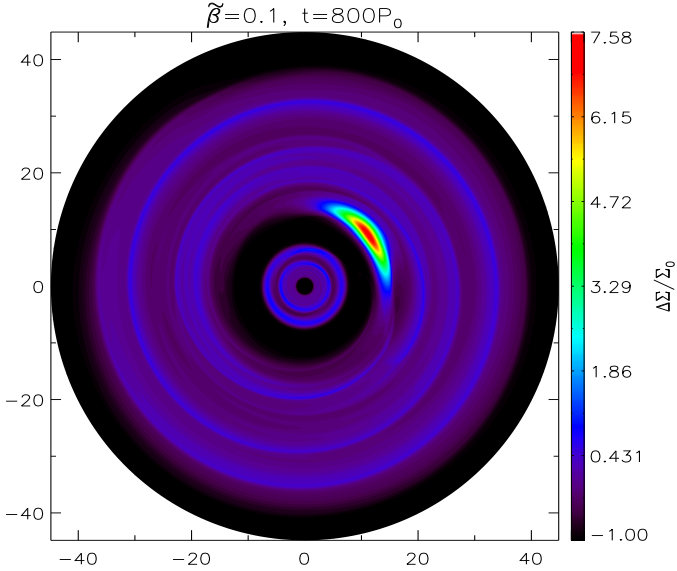}
  \caption{Relative surface density perturbation for the
    $\tilde\beta=0.1$ case during quasi-steady state with a single
    vortex at the outer gap edge. The plot for other values of the cooling time
    $\tilde{\beta}$ are similar. The decrease in the surface
      density near $r\sim 40 r_\mathrm{in}$ arises from mass loss due to the open
      boundary condition imposed at $r_\mathrm{out}$.
    \label{Vortex2D} }
\end{figure}

\begin{figure}
  \includegraphics[width=\linewidth,clip=true,trim=0.5cm
  0cm 0cm 1cm]{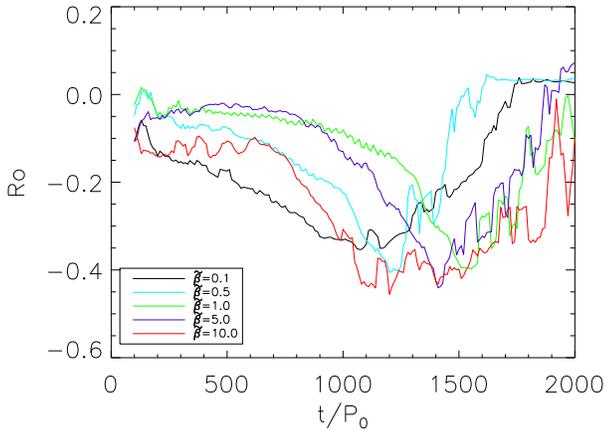}
  \caption{Evolution of Rossby numbers at the centres of the vortices
    formed in discs with different cooling rates. A negative Rossby
    number implies anti-cyclonic motion. Boxcar averaging
    was used to remove contributions from the planet-induced spiral shock.\label{rossbyplot}}    
\end{figure}

In all cases the system remains in a quasi-steady state for
$\gtrsim800P_0$ with a single vortex circulating 
the outer gap edge at the local Keplerian  
frequency. 
Fig. \ref{Vortex2D} shows a typical plot of the relative 
surface density perturbation in this state. During this stage, the 
vortex intensifies. This is better shown in Fig. \ref{rossbyplot} as the
evolution of Rossby numbers measured at the vortex centres.  
The Rossby number increases in magnitude during quasi-steady state,
but the maximum $|Ro|$ is similar for all $\tilde{\beta}$: the vortex
reaches a characteristic value of $Ro\approx-0.35$ for 
$\tilde\beta=0.1$ and $Ro\approx -0.45$ for $\tilde\beta=10$.    


\begin{figure}
  \includegraphics[width=\linewidth,clip=true,trim=0.5cm
  0cm 0cm 1cm]{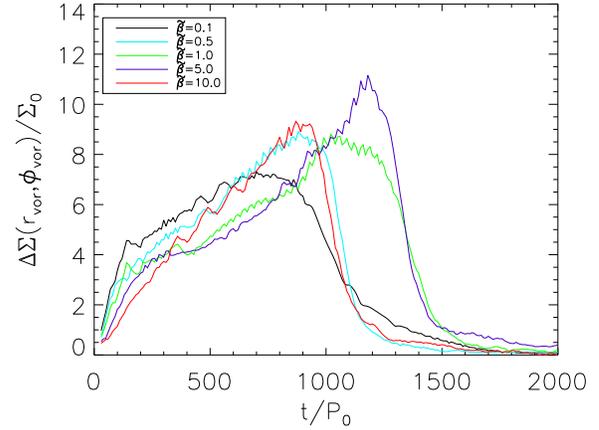}
  \caption{Average value of the relative surface density perturbation
    at vortex centres for various cooling
    times. Initial vortex over-densities decrease with cooling rate while
    vortex growth rates increase with cooling rate.
    \label{overdensity}}     
\end{figure}

We find the vortices become significantly 
over-dense. Fig. \ref{overdensity} plots the surface density perturbation measured at the
vortex centres, showing $\Delta\Sigma/\Sigma_0 \gtrsim 7$ for all
cases of $\tilde\beta$ in quasi-steady state, and 
$\mathrm{max}(\Delta\Sigma/\Sigma_0)\sim 11$ for $\tilde\beta=5$.  
The maximum over-density typically increases with longer cooling
times, despite the vortices are initially weaker at formation with
increasing $\tilde{\beta}$. The large increase in the surface density
is due to vortex growth as there is continuous generation of vorticity by
planet-disc interaction. This is supported by the observation that in
the previous simulations without the planet, the amplitude of
the post-merger vortex does not grow (Fig. 
\ref{planetofflifetimeplot}).  

Fig. \ref{lifetimeplot} shows that the duration of the quasi-steady state
varies with the cooling rate: for 
$\tilde{\beta}=0.1$ and $10$, the vortex amplitude begins to decay around
$t\sim800P_0$, while for $\tilde{\beta}=1,\,5$ the decay begins at 
$t\sim1200P_0$. This non-monotonic dependence suggests that there
exists an optimal cooling rate to maximise the vortex lifetime. We
will discuss this issue further in 
\S\ref{lifetime_discuss}. Notice the decay timescale can be long with
rapid cooling: for $\tilde{\beta}=0.1$ it takes $\sim400P_0$ 
whereas for $\tilde{\beta}=10$ it takes $\sim 100P_0$ for the $m=1$
amplitude to decay significantly after reaching maximum.




\subsection{Additional analysis on vortex decay}
\begin{figure}
  \subfigure{
    \includegraphics[width=0.3\linewidth]{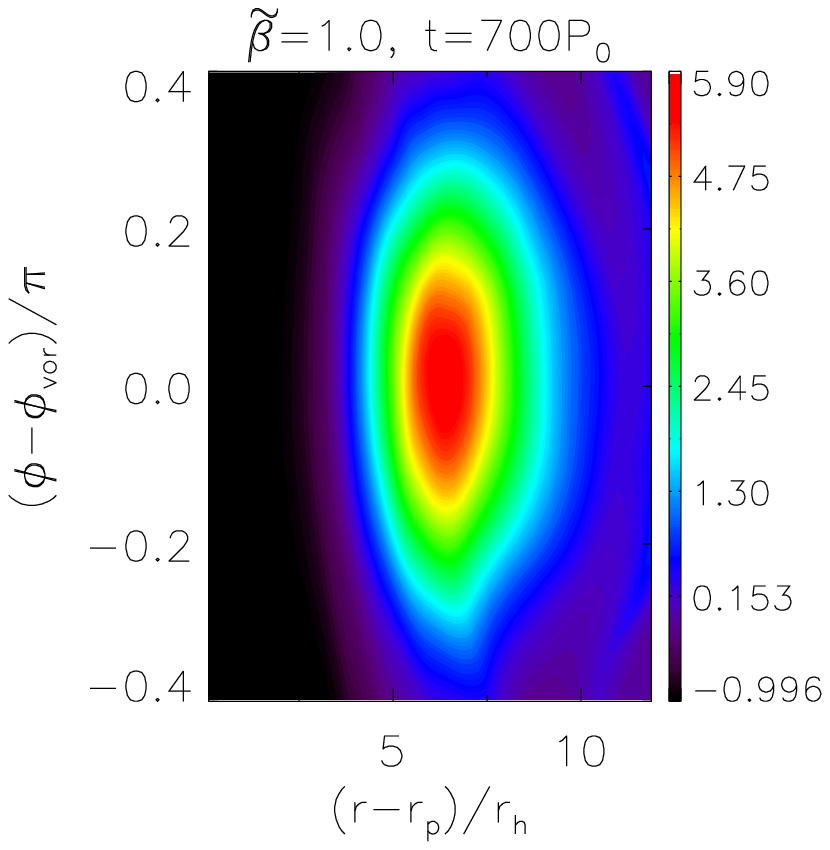}
  }
  \hfill
  \subfigure{
    \includegraphics[width=0.3\linewidth]{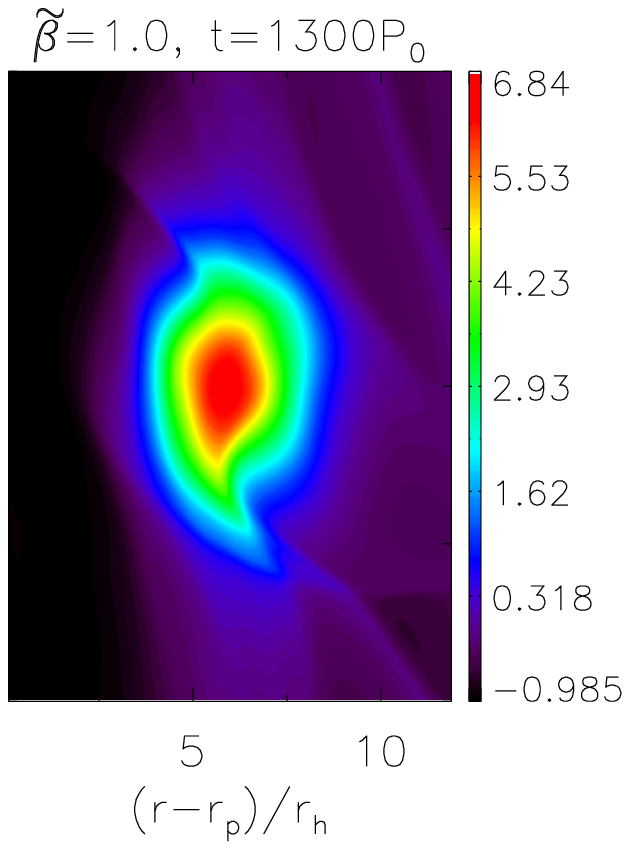}
  }
  \hfill
  \subfigure{
    \includegraphics[width=0.3\linewidth]{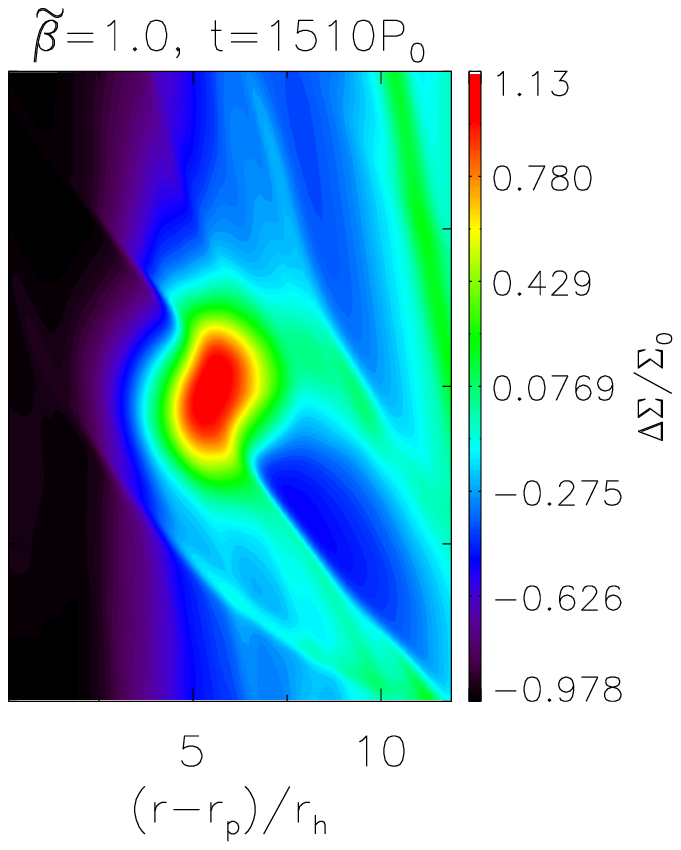}
  } \\[-0.98cm]
  \subfigure{
    \includegraphics[width=0.3\linewidth]{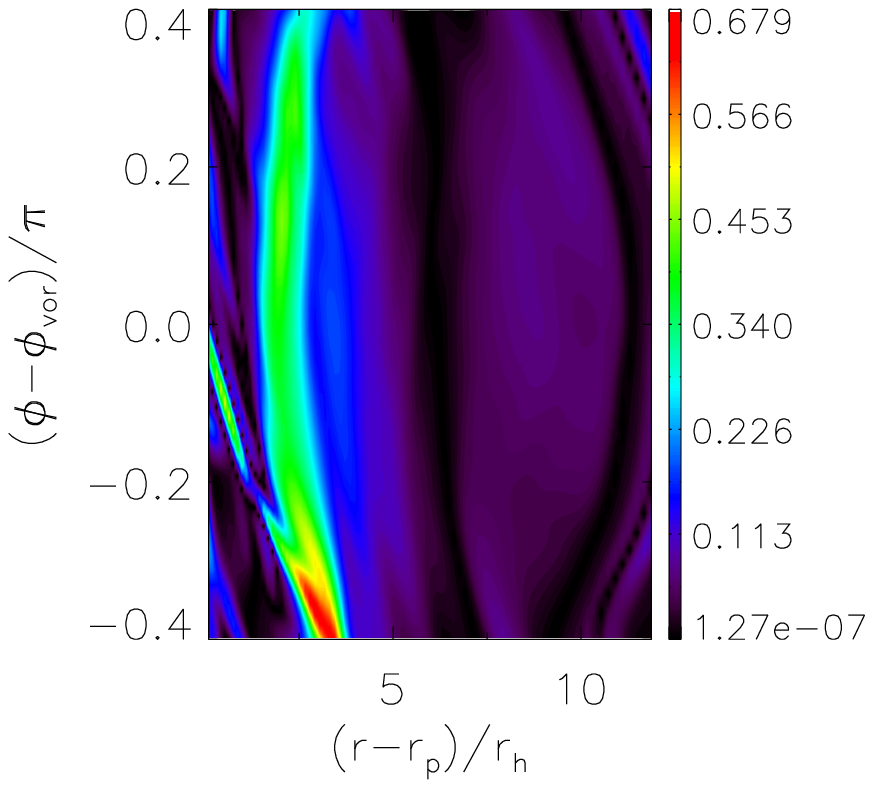}
  }
  \hfill
  \subfigure{
    \includegraphics[width=0.3\linewidth]{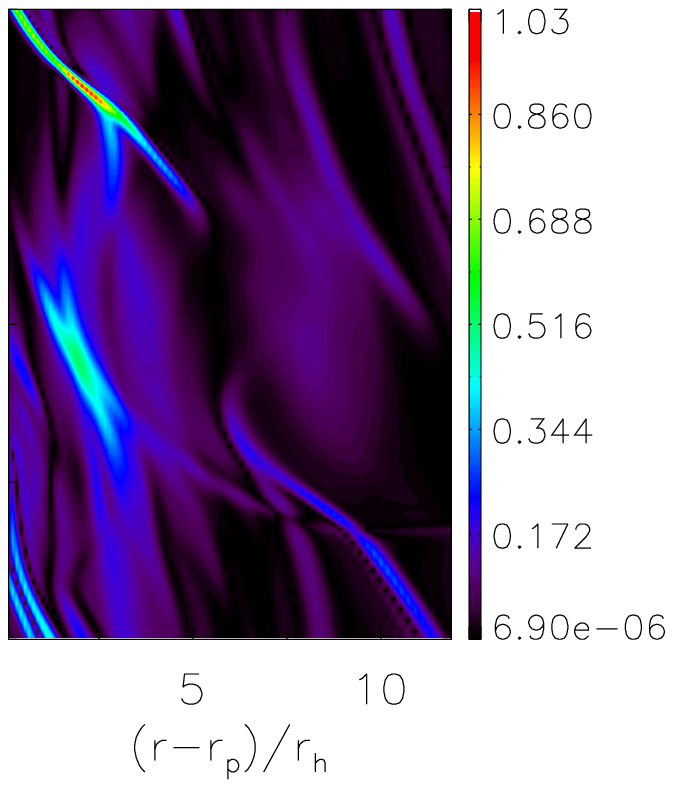}
  }
  \hfill
  \subfigure{
    \includegraphics[width=0.3\linewidth]{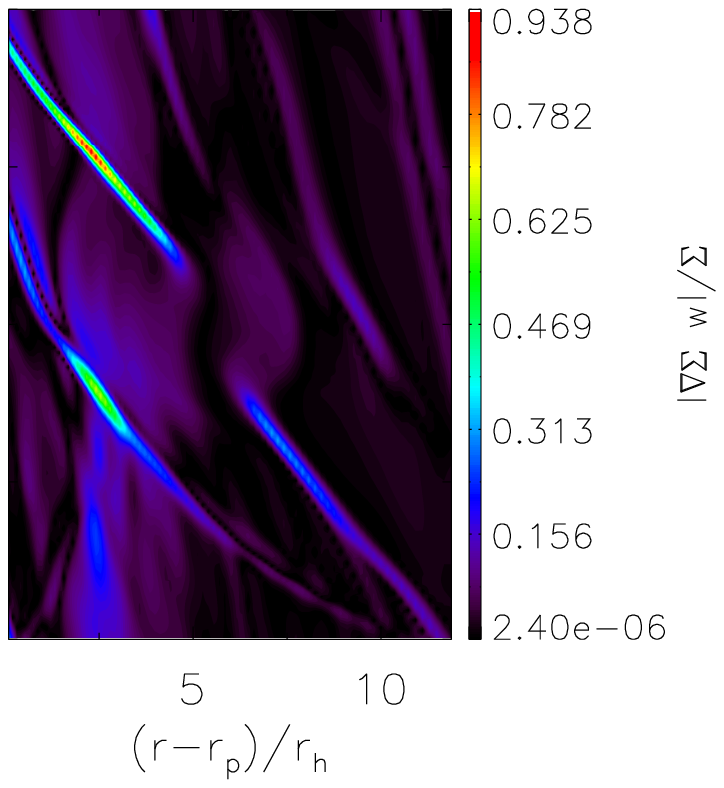}
  } 
  \caption{The vortex in the case with $\tilde{\beta}=1$
    during quasi-steady state (left), start of decay 
    (middle), and just after the decay in the $m=1$ amplitude
    (right). The surface density perturbation
    (top) and the associated surface density gradient (bottom) are
    shown. Wake-like features corresponding to large density gradients
    are found to originate from the vortices during the late phase of 
    their quasi-steady states and into dissipation times.
    This plot is to be considered in conjunction with
    Fig. \ref{lifetimeplot}. 
    \label{shockplot}}
\end{figure}

In this subsection we examine the vortex decay observed in our
simulations in more detail. Fig. \ref{shockplot} show snapshots
of the vortex for the case $\tilde{\beta}=1$. The plots show the surface
density perturbation and the surface density gradient during
quasi-steady state ($t=700P_0$), when the $m=1$ amplitude begins to
decrease ($t=1300P_0$) and just after the rapid amplitude decay
($t=1510P_0$). 

In quasi-steady ($t=700P_0$) the vortex is elongated with a vortex aspect ratio $ \approx 4$
, but becomes more compact approaching a ratio of 2 during its decay ($t=1310P_0$).   
Notice in Fig. \ref{shockplot} the appearance of wakes extending from
either side of the vortex at $t=1300P_0$. These 
wakes correlate with large gradients in surface density (bottom
panel), and are first seen in the later half of the quasi-steady
state. We find the time at which 
the vortex begins to decay coincides with the emergence of these
wakes. 

During the quasi-steady state the vortex 
  orbits at $r\sim r_p+6r_h$. We do not see significant
  vortex migration at this stage, since the vortex is located at a 
  surface density maximum \citep{paardekooper10}. However,
  simultaneous with the appearance of the wakes, we observe the vortex
  begins to migrate inwards to $r\sim r_p+5r_h$.

During quasi-steady state the average value of the surface density gradient
along the wakes is $|w_s\nabla\Sigma/\Sigma| \sim 0.4 $, where 
$w_s\simeq0.1$ (code units) is a typical length scale of 
the surface density variation across the wake.   
Just before the $m=1$ amplitude begins to decrease, we observe this quantity
sharply increases to $ \sim 0.6 $, and remains around this value
until the vortex dies out, at which point the associated Rossby number
begins decreasing to zero. After the 
  vortex reaches small amplitudes ($1\lesssim\Delta\Sigma/\Sigma$), it 
  migrates out to $r\sim r_p+ 6.5 r_h$. 



\begin{figure}
  \includegraphics[width=\linewidth]{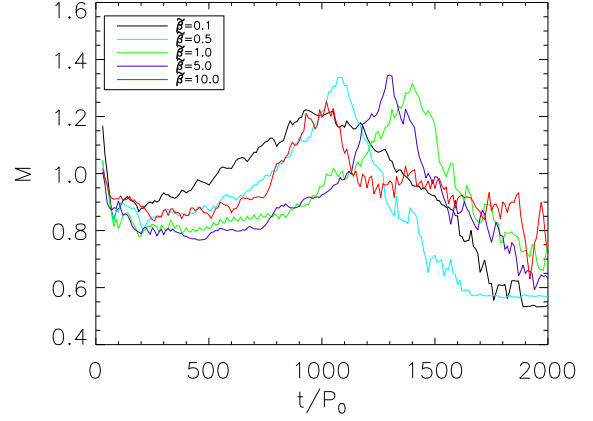}
  \caption{Mach number relative to
    the vortex, averaged over a region within $2H$ of the  
    vortex centre with respect to time. This plot can be compared to the evolution of the
    vortex amplitude shown in Fig. \ref{lifetimeplot}.
    \label{machplot}}
\end{figure}

We also measured large increases in the Mach number near
the vortex as the $m=1$ surface density amplitude reaches maximum and begins to decay. 
Fig. \ref{machplot} plots the Mach number $M=|\bm{v} -
\bm{v}_\mathrm{vor}|/c_s$, where 
$\bm{v}_\mathrm{vor}$ corresponds to the bulk velocity of the vortex
around the disc. Values in Fig. \ref{machplot} have been averaged over
a region within $2H$ of the vortex centre. 
During the quasi-steady state the Mach number increases 
steadily, and for all cases $M$ maximizes about
$\sim 100P_0$ after the start the $m=1$ surface density
amplitude starts to decay. 

Putting the above observations together, we suggest that vortex decay
(in the $m=1$ surface density amplitude) is due to shock formation by
the vortex. When the vortex reaches large amplitude, it begins
to induce shocks in the surrounding fluid, as supported by the
increase in Mach number and the appearance of wakes with large surface
density gradients. The vortex may lose energy through shock
dissipation. In addition, a strong vortex (or shock formation) 
can smooth out the gap structure that originally gave rise to the
RWI, which would oppose vortex growth. We examine this below.

\subsection{Vortex decay and gap structure}  
\begin{figure}
  \includegraphics[width=\linewidth]{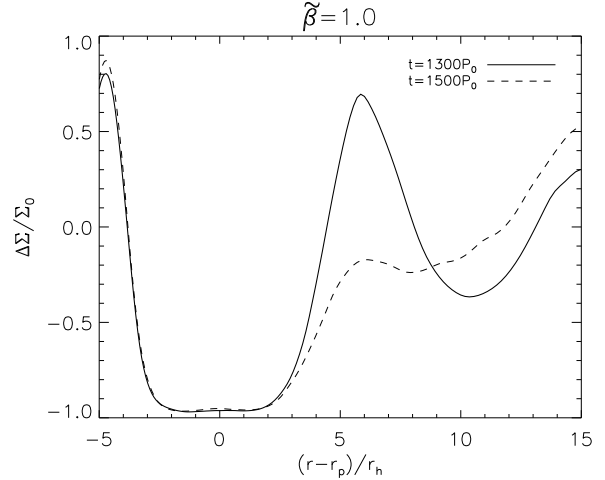}
  \caption{Azimuthally averaged profiles of the relative surface
    density perturbations for  $\tilde\beta=1.0$
    before (solid) and after vortex decay
    (dashed). Alongside drastic reduction in the outer gap maxima, vortex decay
    smooth out the sharp outer gap edge.
    \label{gap_smoothed}} 
\end{figure}

We find vortex decay modifies the gap
structure. Fig. \ref{gap_smoothed} shows the gap profile before and
after vortex decay for the case $\tilde{\beta}=1$. The vortex resides
around the local surface density maximum at the outer gap edge ($r\sim r_p +
6r_h$). We see that after its amplitude has decayed ($t\sim1500P_0$,   
Fig. \ref{lifetimeplot}), this local surface density maximum is also 
smoothed out.

We characterize the smoothness of the outer gap edge with a dimensionless
gap edge gradient parameter
\begin{align}
  \delta\Sigma(t)= \la \frac{\partial \langle \Sigma(t,r)
    \rangle_{\phi}}{\partial r} \cdot \frac{r}{\langle \Sigma(t=0,r)
    \rangle_{\phi}} \ra_{\Delta r}  
\end{align}
where $\Delta r = r\in [r_p,r_p + 6r_h]$ is the radial range of averaging, spanning
from centre of the gap to the radius of the surface density maximum.
A larger $\delta \Sigma$ characterizes a sharper gap edge and  
larger local surface density maxima.   

\begin{figure}
  \includegraphics[width=\linewidth]{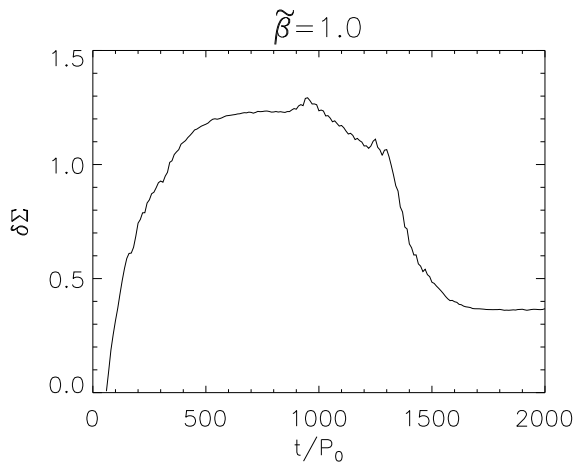}
  \caption{Non-dimensional measure of the surface density gradient at
    the outer gap edge $\tilde\beta=1.0$. During the vortex quasi-steady state
    the gap edge is found to have a large gradient and sharp peak while vortex
    dissipation,
    which occurs at $t\approx10^3P_0$ as seen in Fig. \ref{lifetimeplot},
    works to smooth out the gap edge.
    \label{smoothnessplot}}  
\end{figure}

A plot of the gradient parameter over time
for the $\tilde\beta=1.0$ case is shown in Fig. \ref{smoothnessplot}.
During vortex decay, the outer gap edge is 
drastically smoothed out, changing from a value of $\delta\Sigma=1.2$ during
quasi-steady state to $0.4$ after dissipation. 

This can be interpreted as the vortex providing a viscosity;  
and we measure a typical alpha viscosity $\alpha = O(10^{-2})$
associated with the vortex. This acts against gap-opening
by the planet, and smooths out the outer surface density bump,    
 so the condition for the RWI 
becomes less favourable. In
  order to re-launch the RWI, the surface density bump should
  reform. However, this is difficult as there is no more material
  in the planet's vicinity to clear out (Fig. \ref{gap_smoothed}) to
  form a surface density bump outside the gap. This may explain
why vortices do  not reform again (at least within the simulation
timescale).  After full decay the aspect ratio at the outer gap edge is
  $h\sim0.05$. 

\subsection{Vortex lifetimes as a function of cooling
  rate}\label{lifetime_discuss} 

We now examine vortex lifetimes as function of the imposed cooling
times. For this study, additional simulations with 
  $\tilde\beta=0.01,\,0.25,\,0.75,\,2.5,\,7.5$ were also performed.  
We define the vortex lifetime, $t_{\mathrm{life}}$, as the time
at which the over-density of the vortex
returns to $\Delta \Sigma/\Sigma_0\sim1$ \emph{after} reaching maximum
 (which is on the order of the
initial over-density associated with the gap formation). 

\begin{figure}
  \includegraphics[width=\linewidth]{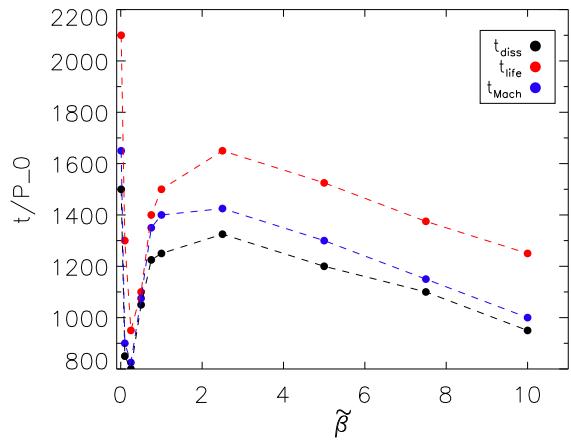}
  \caption{Characteristic timescales associated with vortex evolution, as a function of the cooling
    parameter $\tilde\beta$: time in which the $m=1$
    surface density amplitude begins to decay, $t_\mathrm{diss}$ (black);
    time at which the over-density at the
    vortex centre decreases to $\Delta\Sigma/\Sigma \sim 1$ after
    reaching maximum, $t_\mathrm{life}$ (red);
    and $t_\mathrm{Mach}$ is the time taken
    for the average Mach number around the vortex to maximise
    (blue). \label{betaplot}}  
\end{figure}

We plot $t_{\mathrm{life}}$ with respect to cooling times in
Fig. \ref{betaplot}. We also plot $t_{\mathrm{diss}}$: the time elapsed before the
vortex to begins to dissipate (when the $m=1$ surface density
amplitude begins to decay); and $t_{\mathrm{Mach}}$: the time taken
for the average Mach number around the vortex to maximise. 

For fast cooling rates ($\tilde{\beta}\lesssim 1$), the
vortex lifetime is maximized for $\tilde{\beta}\to0$: we find 
$t_{\mathrm{life}} \approx 2100P_0$ for $\tilde{\beta}=0.01$ and 
decreases to $t_{\mathrm{life}} \approx 950P_0$ for
$\tilde{\beta}=0.25$. Note that for very small $\tilde{\beta}$, 
  there is significant contribution to the overall vortex lifetime due
  to a long decay timescale.
For longer cooling times ($\tilde{\beta}\gtrsim
1$) the vortex lifetimes maximizes at $\tilde\beta=2.5$ with 
$t_{\mathrm{life}}\approx 1650P_0$.  

 We comment here that a locally isothermal simulation, where the
  disc sound-speed is kept constant in time, was also
  performed for comparison. In this case we did not observe
  significant vortex decay within the simulation timescale
 (implying $t_\mathrm{diss}\gtrsim2000P_0$).  We expect the corresponding vortex
 lifetime to exceed that for $\tilde{\beta}=0.01$. Including an energy
 equation with rapid cooling (or setting $\gamma$ close to 
 unity), could still lead to discrepancies with a locally isothermal disc. This is 
 due to the advection-creation of specific entropy within the planet's horseshoe
 region with the former case,  thereby affecting the generalized vortensity 
 and therefore the instability (\S\ref{linear}) and subsequent vortex evolution. 
 Nevertheless, a longer vortex 
 lifetime in a locally isothermal disc would 
 be consistent with the above trend of increasing vortex lifetimes as
 $\tilde{\beta}\to0$. 

In the previous section, we observed that vortices began to decay when it starts to
induce shocks.  We thus suggest that the time needed for the vortex to grow to 
sufficient amplitude to induce shocks in the surrounding fluid,   
which may be considered as the duration of the quasi-steady state or
$t_\mathrm{diss}$, is an important contribution to the overall vortex
lifetime.
We discuss below some competing factors that may result in a 
non-monotonic dependence of $t_\mathrm{diss}$ on the cooling rate.  

\subsubsection{Factors that lengthen vortex lifetimes}
It has been shown that the amplitude at which the RWI saturates 
increases with the growth rate of the linear instability  
\citep{meheut2013}. Our `planet-off' simulations yield slower growth
rates with increasing cooling times, which suggest weaker vortices are
formed initially with increasing $\tilde{\beta}$. This is consistent
with the present simulations: at the beginning of the 
quasi-steady  state ($t\sim100P_0$) we find the over-density at the
vortex centre is $\Delta\Sigma/\Sigma_0=2.5$ for $\tilde\beta=0.1$ and
$\Delta\Sigma/\Sigma_0=1.48$ for $\tilde\beta=10$.

The growth of the post-merger single vortex is mediated by disc-planet
interaction. However, gap-opening becomes more difficult in a hotter 
disc, and we find the generalised vortensity 
profiles are smoother with increasing $\tilde{\beta}$. This opposes
the RWI. 
Furthermore, the vortex should reach larger amplitudes to induce
shocks on account of the increased sound-speed.  

These considerations suggest, with increased cooling times, 
it takes longer for the post-merger vortex grow to sufficient
amplitude to induce shocks and dissipate. This factor contributes to a
longer quasi-steady state with increasing $\tilde{\beta}$.       


\subsubsection{Factors that shorten vortex lifetimes} 
Notice in Fig. \ref{lifetimeplot} and Fig. \ref{overdensity}, the vortex 
growth during the quasi-steady state is actually faster for $\tilde{\beta}=10$
than for $\tilde{\beta}=5$. 
For example, at $t\sim 500P_0$ the vortex with    
$\tilde{\beta}=10$ has a larger 
amplitude than for $\tilde{\beta}=5$. This is also reflected in Fig. \ref{machplot}, where the
Mach number reaches its maximum value sooner for $\tilde{\beta}=10$
than for $\tilde{\beta}=5$.  

This observation is consistent with the RWI being favoured by 
higher temperatures \citep{li00,lin12c} through the perturbations 
(as opposed to its effect through the set up of the gap profile
discussed previously),
which corresponds to longer cooling times in our case. 
While our `planet-off' simulations indicate this is unimportant for
the linear instability, it may have contributed  
significantly to the vortex growth during quasi-steady state at very
long cooling times (e.g. $\tilde{\beta}=10$). This effect shortens the
vortex lifetime by allowing it to grow faster and induce shocks
sooner.  

%

\section{Summary and discussion}\label{summary}
In this paper, we have carried out numerical simulations of
non-isothermal disc-planet interaction.  
Our simulations were customized to examine the effect of a finite 
cooling time on the stability of gaps  
opened by giant planets to the so-called vortex or Rossby wave
instability. To do so, we  
included an energy equation with a cooling term that restores the 
disc temperature to its initial profile on a characteristic timescale
$t_c$. We studied the evolution of the gap stability as a function of 
$t_c$. This is a natural extension to previous studies of on gap
stability, which employ locally or strictly isothermal equations of
state. We considered the inviscid limit which favors the RWI
  \citep{li09,fu14} and avoids complications from viscous
  heating other that shock heating. However, this means that the vortex lifetimes observed in
  our simulations are likely longer than in realistic discs with
  non-zero physical viscosity.    	

We considered two types of numerical experiments. We first used
disc-planet interaction to self-consistently set up gap profiles,
which were then perturbed and evolved without further the influence of
the planet potential. This procedure isolates the effect of cooling on
gap stability through the set up of the initial gap profile. We find
that as the cooling time $t_c$ is increased, the gaps became more
stable, with lower growth rates of non-axisymmetric modes and the
dominant azimuthal wavenumber also decreases. This is consistent with
the notion that increasing $t_c$ leads to higher temperatures or
equivalently the disc aspect ratio $h$,
which opposes gap-opening by the planet. This means that the gaps
opened by the planet in a disc with longer $t_c$ are smoother and
therefore  more stable to the RWI. 

In the second set of calculations, we included the planet potential
throughout the simulations  and examined the long-term evolution of
the gap-edge vortex that develops from the RWI. The vortex reaches 
a quasi-steady state lasting $O(10^3)$ orbits. Unlike the `planet-off'
simulations, in which vortices decay after linear growth and merging,
we find that with the planet potential kept on, the vortex amplitude
grows during this quasi-steady state, during which no vortex migration is
  observed, until it begins to induce
shocks, after which the vortex amplitude begins to decay.   

For our main simulations with $\tilde{\beta}\geq 0.1$, the duration of the quasi-steady state increases with
increasing cooling timescales until a critical value, beyond which this
quasi-steady state shortens again.  We find the timescale for the vortex to
decay after reaching maximum amplitude can be long for small
$\tilde{\beta}$, which contributes to a long overall vortex lifetime
with rapid cooling. We do observe vortex migration during its
decay, which may influence this decay timescale. 
  

We suggest a non-monotonic dependence of the quasi-steady state 
on the cooling timescale $\tilde{\beta}$ can
be attributed to the time required for the vortex to grow to 
sufficient amplitude to induce shocks in the surrounding fluid,
thereby losing energy and also smooth out the gap edge.   

For short cooling timescales, the planet is able to open a
deeper gap which favours the RWI, leading to stronger 
vortices. For long cooling timescales, we find the vortex
grows faster during the quasi-steady state. In accordance with
previous stability calculations \citep{li00}, we 
suggest the latter is due to the RWI being favoured with increasing
disc temperature, and that this effect overcomes weaker
gap-opening for sufficiently long cooling times. 
These competing factors imply
for both short and long cooling timescales, the vortex reaches
its maximum amplitude, shock, and begins to decay, sooner than
intermediate cooling timescales. 

(However, for very rapid cooling, e.g. $\tilde{\beta}=0.01$, the quasi-steady state is also
  quite long. This suggests that the above effects themselves do not
  have a simple dependence on the cooling timescale when considering
  $\tilde{\beta}\to0$ and/or that other factors become important in
  this limit. This should be investigated in future works.)


We remark that a non-monotonic dependence of the vortex lifetime was
also reported by \cite{fu14}, who performed  locally isothermal disc-planet
simulations with different values of the   
disc aspect ratio. In their simulations the optimum aspect ratio is
$h=0.06$. In our simulations, $h$ is a dynamical
variable, but by analyzing the region where the vortex is located
($r-r_p\in[2,10]r_h$), 
we find for a dimensionless cooing timescale of $\tilde\beta=2.5$, which has
the longest vortex lifetime in the presence of moderate cooling, that  
$h\approx0.058$ on average.  
Our result is consistent with \cite{fu14}.


\subsection{Caveats and outlooks}
There are several outstanding issues that needs to be addressed
in future work: 

\emph{Convergence.} Although lower resolution simulations performed in
the early stages of this project gave similar results  
(most importantly, the non-monotonic dependence of vortex lifetimes on
the cooling timescale), we did find the lower resolution 
typically yield longer vortex lifetimes than that reported in this
paper. This could be due to weaker RWI with low resolution.  
It will be necessary to perform even higher resolution  simulations
in order to obtain quantitatively converged vortex lifetimes.

  \emph{Orbital migration.} We have held the planet fixed on a circular
  orbit. However, gap-edge vortices are known to exert
  significant, oscillatory torques on the planet \citep{li09}
  which can lead to complex orbital migration. This will likely affect
  vortex lifetimes as it may alter the planet-vortex separation, as
  well as leading to direct vortex-planet interactions
  \citep{lin10,ataiee14}. Thus, 
  future simulations should allow the planet to freely
  migrate. Similarly, the role of vortex migration on its lifetime
  should be clarified.

  \emph{Cooling model.} Our prescription for the disc 
  heating/cooling is convenient to probe the full range of
  thermodynamic response of the disc. However, in order to calculate 
  vortex lifetimes in actual 
  protoplanetary discs, an improved thermodynamics treatment,
  e.g. radiative cooling based on realistic disc temperature, density,
  opacity models etc., should be used in future work.   

\emph{Self-gravity.} We have ignored disc self-gravity in this
study. Based on linear calculations, \cite{lovelace13} concluded
self-gravity to be important for the RWI when the Toomre parameter $Q<O(1/h)$, or
$Q\lesssim 20$ for $h\sim0.05$, as was typically considered in this
work. This suggests that self-gravity may affect vortex lifetimes even
when $Q$ is not small. In particular, given that
we observe vortices can reach significant over-densities (up to almost
an order of magnitude), it will be important to include disc
self-gravity in the future. 
 
\emph{Three-dimensional (3D) effects.} A vortex in a 3D disc may be
subject to secondary instabilities that destroy them
\citep{lesur09,railton14}. This may be an important factor in
determining gap-edge vortex lifetimes in realistic discs. For example,
if these secondary instabilities sets in before the vortex grows to
sufficient amplitude to shock, then the dependence of the vortex
lifetime on the cooling timescale will be its effect through the 3D
instability (as opposed to the effect on the RWI itself, which is a 2D
instability). This problem needs to be clarified with full 3D
disc-planet simulations.  


\section*{Acknowledgments}
This project was initiated at the Canadian Institute for Theoretical
Astrophysics (CITA) 2014 summer student programme. The authors
  thank the anonymous referee for an insightful report. 
Computations were performed on the GPC supercomputer at the
SciNet HPC Consortium. SciNet is funded by: the Canada Foundation for
Innovation under the auspices of Compute Canada; the Government of
Ontario; Ontario Research Fund - Research Excellence; and the
University of Toronto.   

\bibliographystyle{mn2e}
\bibliography{ref}

\appendix

\end{document}